\documentclass{article} 
\usepackage{longcat_style,times}

\usepackage{amsmath,amsfonts,bm}

\def\eqref#1{equation~\ref{#1}}

\def\1{\bm{1}}

\DeclareMathAlphabet{\mathsfit}{\encodingdefault}{\sfdefault}{m}{sl}
\SetMathAlphabet{\mathsfit}{bold}{\encodingdefault}{\sfdefault}{bx}{n}

\usepackage{hyperref}
\usepackage{url}

\usepackage{graphicx}
\usepackage{balance}  
\usepackage{amsmath}
\usepackage{algpseudocode}
\usepackage{latexsym}
\usepackage{multirow}
\usepackage{multicol}
\usepackage{color}
\usepackage{dashrule}
\usepackage{extarrows}
\usepackage{dsfont}
\usepackage{centernot}
\usepackage{hyperref}
\usepackage{natbib}
\usepackage{fancybox}
\usepackage[utf8]{inputenc} 
\usepackage[T1]{fontenc}    
\usepackage{hyperref}       
\usepackage{url}            
\usepackage{booktabs}       
\usepackage{amsfonts}       
\usepackage{amsmath}
\usepackage{nicefrac}       
\usepackage{microtype}      
\usepackage{xcolor} 
\usepackage{amsmath}
\usepackage{lipsum}
\usepackage{colortbl}
\usepackage{subcaption}
\usepackage{tabularx}
\usepackage{makecell}
\usepackage{wrapfig}
\usepackage{bm}
\usepackage{multirow}
\usepackage{xcolor} 
\usepackage{bbm}
\usepackage{mdframed}
\usepackage{enumitem}
\usepackage{xspace}

\usepackage{lineno}

\usepackage{xurl}
\usepackage{subcaption}
\usepackage{amsmath}
\usepackage{multirow}
\usepackage{makecell}
\usepackage{booktabs}
\usepackage{bm}

\usepackage[T1]{fontenc}

\usepackage[utf8]{inputenc}

\usepackage{microtype}

\usepackage{inconsolata}

\usepackage{graphicx}

\usepackage{amsmath}
\usepackage{amssymb}
\usepackage{booktabs}
\usepackage{multirow}
\usepackage{makecell}
\usepackage{tcolorbox}
\usepackage{listings}
\usepackage{xcolor}
\usepackage[table]{xcolor} 
\usepackage{graphicx}
\usepackage{amsmath}

\definecolor{promptbg}{RGB}{248,248,248}
\definecolor{promptframe}{RGB}{210,210,210}

\lstdefinestyle{promptstyle}{
  basicstyle=\ttfamily\small,
  columns=fullflexible,
  breaklines=true,
  breakatwhitespace=true,
  keepspaces=true,
  showstringspaces=false,
  frame=none,
  tabsize=2,
  gobble=0,
  breakindent=0pt,         
  breakautoindent=false,   
  postbreak=\mbox{}        
}

\newtcolorbox{promptbox}[1][Prompt]{
  colback=promptbg,
  colframe=promptframe,
  title=\textbf{#1},
  boxrule=0.6pt,
  arc=1mm,
  coltitle=black,        
  colbacktitle=gray!25,  
  boxsep=4pt,            
  left=4pt, right=4pt, top=4pt, bottom=4pt 
}
\definecolor{darkblue}{rgb}{0, 0, 0.5}
\hypersetup{colorlinks=true, citecolor=darkblue, linkcolor=darkblue, urlcolor=darkblue}

\usepackage{listings}
\usepackage{xcolor}
\usepackage[ruled,vlined]{algorithm2e}
\usepackage{caption}
\usepackage{float} 
\usepackage{arydshln}
\usepackage{wrapfig} 
\usepackage{tcolorbox}
\tcbuselibrary{theorems}
\tcbuselibrary{most}
\usepackage[dvipsnames]{xcolor}

\definecolor{customTeal}{RGB}{0, 128, 128} 
\definecolor{emphasisColor}{RGB}{255, 0, 0} 
\definecolor{oursBlue}{RGB}{51,202,246}
\newcommand{\name}{\textsc{DIFFA-2}\xspace}
\definecolor{blue1}{HTML}{508AB2}
\definecolor{green2}{HTML}{BFF6BA}

\newtcbtheorem[]{prompt}{Prompt}%
{colback=SeaGreen!10!CornflowerBlue!10,
 colframe=RoyalPurple!55!Aquamarine!100!,
 fonttitle=\bfseries,
 left=.02in, right=.02in, bottom=.02in, top=.02in,
 before upper={\linespread{1.5}\selectfont}}{prompt}

\renewcommand{\headeright}{\raisebox{-0.2\height}{\includegraphics[height=2.2em]{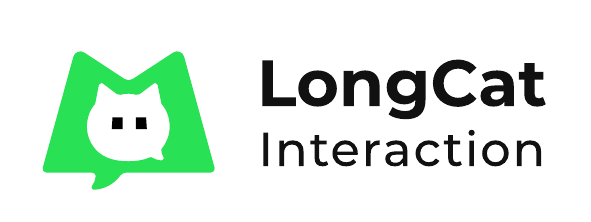}}}
\renewcommand{\shorttitle}{\name: A Practical Diffusion Large Language Model for General Audio Understanding}

\definecolor{darkblue}{rgb}{0, 0, 0.5}
\hypersetup{colorlinks=true, citecolor=darkblue, linkcolor=darkblue, urlcolor=darkblue}

\makeatletter
\renewcommand{\@maketitle}{%
  \vbox{%
    \hsize\textwidth
    \linewidth\hsize
    \vskip -0.5in
    \noindent
    \begin{minipage}{0.2\textwidth}
      \includegraphics[width=\linewidth]{pic/Logo.pdf}
    \end{minipage}%
    \\
    \rule{\linewidth}{1pt}
    \hspace{0.05\textwidth}%
    \begin{minipage}{0.8\textwidth}
    \end{minipage}

    \centering
    {\LARGE \bfseries\@title\par}
    \vskip 0.1in  
    \def\And{%
      \end{tabular}\hfil\linebreak[0]\hfil%
      \begin{tabular}[t]{c}\bf\rule{\z@}{24\p@}\ignorespaces%
    }
    \def\AND{%
      \end{tabular}\hfil\linebreak[4]\hfil%
      \begin{tabular}[t]{c}\bf\rule{\z@}{24\p@}\ignorespaces%
    }
    \begin{tabular}[t]{c}\bf\rule{\z@}{24\p@}\@author\end{tabular}%
  \vskip 0.05in 
  }
}
\makeatother

\title{\name: A Practical Diffusion Large Language Model for
General Audio Understanding \\}

\makeatletter
\def\@fnsymbol#1{\ensuremath{\ifcase#1\or \dagger\or \ddagger\or
   \mathsection\or \mathparagraph\or \|\or **\or \dagger\dagger
   \or \ddagger\ddagger \else\@ctrerr\fi}}
\makeatother

\author{
\begin{tabular}{c}
\textbf{Jiaming Zhou}$^{1,2}$\thanks{This work was done during an internship at Meituan.}
\quad
\textbf{Xuxin Cheng}$^{2}$
\quad
\textbf{Shiwan Zhao}$^{1}$
\quad
\textbf{Yuhang Jia}$^{1,2 \dagger}$
\quad
\textbf{Cao Liu}$^{2}$ \\[1ex]
\textbf{Ke Zeng}$^{2}$
\quad
\textbf{Xunliang Cai}$^{2 \ddagger}$ 
\quad
\textbf{Yong Qin}$^{1}$\thanks{Xunliang Cai and Yong Qin are corresponding authors. Correspondence to: qinyong@nankai.edu.cn} \\[1ex]
\quad
\normalfont $^1$College of Computer Science, Nankai University
\quad
$^2$Meituan LongCat Interaction Team\\[1ex]
\normalfont \texttt{\url{https://github.com/NKU-HLT/DIFFA}}
\end{tabular}
}

\begin{document}

\maketitle

\begin{abstract}
Autoregressive (AR) large audio language models (LALMs) such as Qwen-2.5-Omni have achieved strong performance on audio understanding and interaction, but scaling them remains costly in data and computation, and strictly sequential decoding limits inference efficiency. 
Diffusion large language models (dLLMs) have recently been shown to make effective use of limited training data, and prior work on DIFFA indicates that replacing an AR backbone with a diffusion counterpart can substantially improve audio understanding under matched settings, albeit at a proof-of-concept scale without large-scale instruction tuning, preference alignment, or practical decoding schemes.
We introduce \textbf{DIFFA-2}, a practical diffusion-based LALM for general audio understanding. DIFFA-2 upgrades the speech encoder, employs dual semantic and acoustic adapters, and is trained with a four-stage curriculum that combines semantic and acoustic alignment, large-scale supervised fine-tuning, and variance-reduced preference optimization, using only fully open-source corpora. Experiments on MMSU, MMAU, and MMAR show that DIFFA-2 consistently improves over DIFFA and is competitive to strong AR LALMs under practical training budgets, supporting diffusion-based modeling is a  viable backbone for large-scale audio understanding.

\end{abstract}

\section{Introduction}

Diffusion large language models (dLLMs)~\citep{llada,llada1.5} have recently emerged as a promising alternative to conventional autoregressive (AR) decoders. Instead of generating tokens strictly left-to-right, dLLMs perform iterative denoising over partially masked sequences, enabling any-order token modeling and naturally supporting parallel token updates during decoding. Recent study~\citep{ni2025diffusion} in the text domain further shows that dLLMs can act as strong \emph{data learners}: when the amount of unique training data is constrained, they continue to improve and can even surpass AR models by leveraging super-dense compute and implicit Monte Carlo-style data augmentation. These properties are particularly appealing for audio understanding, where high-quality audio--text supervision across speech, sound, and music is far more expensive to collect than text-only data, and where the latency of strictly sequential AR decoding becomes a bottleneck for long-form and interactive applications.

Despite these advantages, state-of-the-art large audio language models (LALMs) are still predominantly AR-based: systems such as Qwen-3-Omni~\citep{Qwen3-Omni}, Qwen-2.5-Omni~\citep{Qwen2.5-0mni}, and Kimi-Audio~\citep{Kimi-audio} couple powerful speech encoders with AR LLMs and achieve strong results on a wide range of audio understanding and dialogue benchmarks. This raises a natural question: \emph{can dLLMs be turned into competitive and practical audio backbones that match these AR LALMs under realistic data and latency budgets?} Initial evidence is encouraging. A recent work on diffusion-based large audio-language model (DIFFA)~\citep{diffa} compares an 8B AR backbone with its diffusion counterpart under matched data, adapter design, and training recipe, and finds substantial gains on audio understanding benchmarks such as MMAU~\citep{sakshi2025mmau} and MMSU~\citep{wang2025mmsu} after simply replacing the AR backbone. This suggests that the generative paradigm itself can strongly influence audio performance and that dLLMs have significant potential as audio backbones. However, that study remains largely a proof of concept: the model is trained mainly on speech-centric supervision with a relatively small Whisper encoder~\citep{whisper}, keeps the diffusion backbone frozen, and does not exploit large-scale instruction data, preference-based objectives, or practical inference acceleration. It therefore does not answer whether dLLMs can be scaled into \emph{strong and practical} LALMs that reliably compete with state-of-the-art AR models.

In this paper, we present \textbf{DIFFA-2}, a substantially strengthened diffusion-based large audio language framework that addresses this question. DIFFA-2 aims to turn dLLMs from a proof-of-concept backbone into a competitive and practical audio model through comprehensive semantic--acoustic alignment and scalable training. Concretely, it adopts a four-stage training strategy that progressively aligns audio representations, incorporates large-scale supervised fine-tuning, and applies variance-reduced preference optimization (VRPO)~\citep{llada1.5}, and combines this with factor-based parallel decoding~\citep{fastdllm} at inference time to enable practical audio-based interaction without relying on AR decoding. Our contributions are as follows:
\begin{itemize}[itemsep=2pt,topsep=0pt,parsep=0pt]
    \item We present \textbf{DIFFA-2}, a strengthened diffusion-based large audio language model with improved acoustic modeling and semantic--acoustic alignment for unified understanding of speech, sound, and music.

    \item We introduce a progressive four-stage training curriculum that combines semantic and acoustic alignment, large-scale supervised fine-tuning, and preference-based reinforcement learning using fully open-source data, together with factor-based parallel decoding for practical diffusion inference.

    \item With only 11{,}000 hours of automatic speech recognition (ASR) data and 3{,}767 hours of supervised fine-tuning data, DIFFA-2 updates only about 1.1\% of parameters and achieves strong performance on audio understanding benchmarks including MMSU, MMAU, and MMAR~\citep{ma2025mmar}, remaining highly competitive with strong AR-based LALMs on these benchmarks.

    \item We open-source both the training and inference pipeline to facilitate future research on dLLM-based audio models.
\end{itemize}

\section{Preliminaries}

LLaDA~\citep{llada} is a non-autoregressive language modeling framework based on a discrete random masking process.
Instead of factorizing the sequence likelihood in a left-to-right manner, LLaDA introduces a stochastic corruption mechanism and trains a mask predictor to approximate the reverse denoising process.
This design allows the model to leverage bidirectional context and enables parallel token prediction during inference.

Formally, given a clean target sequence $x_0 = (x_0^1, \dots, x_0^L)$, LLaDA defines a forward masking process that independently replaces each token with a special mask symbol $\mathrm{M}$ with probability $t \in (0,1]$, resulting in a corrupted sequence $x_t$.
The mask predictor $p_\theta(x_0 \mid x_t)$, parameterized by a standard Transformer decoder, is trained to reconstruct the original tokens at masked positions.
The pre-training objective is given by
\begin{equation}
\label{eq:pretrain-obj}
\begin{aligned}
\mathcal{L}(\theta) \triangleq 
- \mathbb{E}_{t, x_0, x_t}
\left[
\frac{1}{t}
\sum_{i=1}^{L}
\mathbb{I}[x_t^i = \mathrm{M}]
\log p_\theta(x_0^i \mid x_t)
\right],
\end{aligned}
\end{equation}
where $L$ denotes the sequence length.
This objective yields a tractable upper bound on the negative log-likelihood~\citep{shi2024simplified,ou2025your}, while avoiding autoregressive factorization.

Supervised fine-tuning (SFT) under the LLaDA framework follows the same corruption--reconstruction principle.
Given a prompt--response pair $(p_0, r_0)$, only the response sequence is subject to random masking, producing a corrupted response $r_t$, while the prompt remains fully observed.
The SFT objective is defined as
\begin{equation}
\label{eq:sft-objective}
- \mathbb{E}_{t, p_0, r_0, r_t}
\left[
\frac{1}{t}
\sum_{i=1}^{L'}
\mathbb{I}[r_t^i = \mathrm{M}]
\log p_\theta(r_0^i \mid p_0, r_t)
\right],
\end{equation}
where $L'$ denotes the response length.

At inference time, LLaDA performs generation through an iterative decoding procedure.
Starting from a fully masked response sequence, the model predicts token values at masked positions and selectively re-applies masks to low-confidence tokens.
By repeating this denoising process for a fixed number of steps, the model gradually refines its predictions and produces the final output sequence.

\begin{figure*}[!t]
\centering
\includegraphics[width=0.97\textwidth]{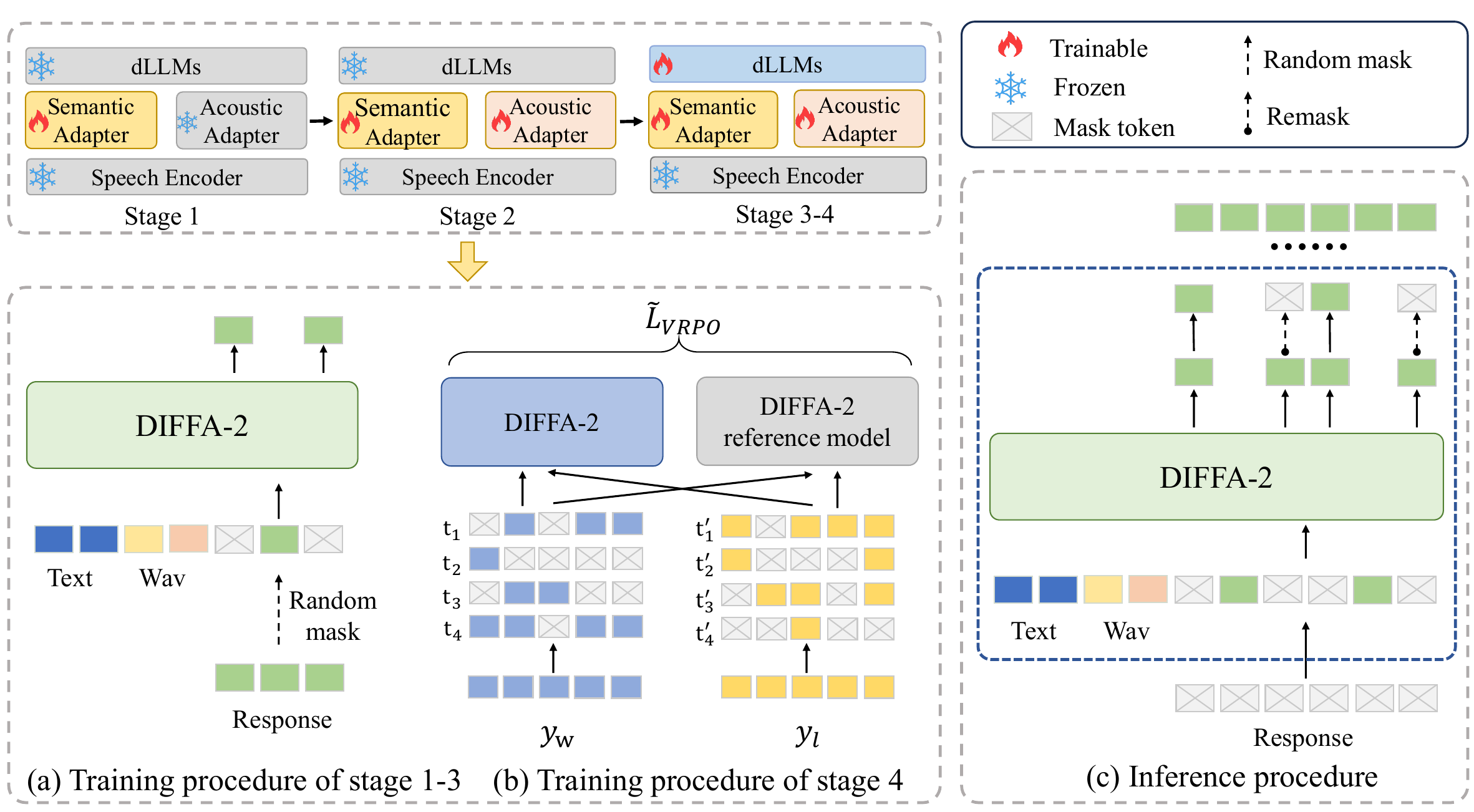}
\caption{Overview of DIFFA-2, including the dual-adapter architecture, multi-stage training pipeline (Stages 1–4), and iterative diffusion-based inference for general audio understanding.}
\renewcommand{\arraystretch}{1.1}
\label{pic:train}
\end{figure*}

\section{Methods}
\label{sec:method}
\subsection{Model Architecture}
\label{sec:arch}

DIFFA-2 follows the overall framework of DIFFA, but substantially strengthens acoustic representation and cross-modal alignment. The model consists of three main components: a frozen Whisper-Large-V3 encoder, a dual-adapter audio interface, and a dLLM backbone. 
The \textit{semantic adapter} consists of a two-layer convolution subsampling module followed by a two-layer linear projection, reducing the temporal resolution from 50~Hz to 12.5~Hz and aligning temporally aggregated audio features with textual semantics. The \textit{acoustic adapter} is implemented as a two-layer Q-former~\citep{li2023blip} with 64 trainable query vectors attending to intermediate encoder states, enabling effective capture of paralinguistic and non-linguistic acoustic cues, such as prosody, emotion-related patterns, environmental sounds, and music.
The key design idea is to expose the backbone to two complementary views of the audio signal for speech, sound, and music: a content-oriented stream that is temporally aligned with textual semantics, and a compact acoustic summary stream that highlights prosodic, stylistic, and non-linguistic cues, while keeping the diffusion backbone lightweight to train.

\subsection{Training Data Overview}
\label{sec:data-overview}

Our training pipeline follows a progressive curriculum corresponding to the four stages detailed in Sec.~\ref{sec:training-strategy}. Broadly, we utilize large-scale transcription data for semantic alignment, diverse audio-centric instruction data for acoustic enrichment, and curated preference pairs for reinforcement-style alignment. Detailed data construction and statistics are provided in Appendix~\ref{appendix:data-details}.

\paragraph{Transcription Data for Semantic Alignment.} For Stage~1, we leverage major ASR corpora, including LibriSpeech~\citep{panayotov2015librispeech} and GigaSpeech~\citep{chen2021gigaspeech}, by framing speech recognition as an instruction-following task. We utilize Qwen3-32B~\citep{yang2025qwen3} to generate 25 distinct instruction templates, which are applied to original transcriptions. 

\paragraph{SFT Data.} To move beyond pure speech recognition and enrich the model's acoustic and paralinguistic understanding (Stages~2 and~3), we construct four complementary types of audio question answering (AQA) data: (i) Caption-grounded AQA: Using multi-domain audio and speech corpora with existing captions or paralinguistic annotations, we prompt a strong LLM to synthesize diverse and high-quality grounded answers. This ensures broad yet controlled coverage of speech, environmental sounds, and music. (ii) Direct audio QA via text-to-speech (TTS), where text QA pairs from general text datasets are converted into speech via TTS, covering simple and complex queries. (iii) Multiple-choice AQA: This subset, derived from existing benchmarks, emphasizes fine-grained discrimination and objective evaluation of specific audio attributes. (iv) ASR Subset: a subset from ASR data. 

\paragraph{Preference Alignment Data.} In Stage~4 (preference optimization), we construct high-quality preference triplets consisting of an audio input, a question, and a reference answer. We prompt an LLM to generate "rejected" responses that are fluent and superficially plausible but contain subtle audio-related factual errors (e.g., incorrect gender, rhythm, or sound events). Only pairs where the reference is unambiguously superior are retained. These chosen--rejected pairs are utilized for VRPO to sharpen the model’s sensitivity to nuanced paralinguistic cues and complex audio reasoning.

\subsection{Progressive Four-Stage Training}
\label{sec:training-strategy}

To fully exploit the dual-adapter architecture and the heterogeneous training data, we adopt a progressive four-stage training curriculum. Throughout all stages, the Whisper-large-v3 encoder is kept frozen, and supervised objectives operate only on the adapters and the diffusion backbone.

\paragraph{Stage 1: Semantic alignment on ASR.}
In the first stage, we freeze the dLLM backbone and train only the semantic adapter under an ASR-style objective. The goal is to align it with the semantic space, so that its outputs can be seamlessly consumed by the diffusion backbone.

\paragraph{Stage 2: Joint semantic--acoustic alignment.}
In stage~2, we still keep the diffusion backbone frozen, but jointly align both semantic and acoustic adapters with synthesized SFT data that explicitly incorporates paralinguistic and acoustic cues, thereby enhancing both semantic and acoustic understanding in speech, audio, and music.

\paragraph{Stage 3: Unfreezing the diffusion backbone with LoRA.}
In this stage, we employ Low-Rank Adaptation (LoRA)~\citep{hu2022lora} to fine-tune the diffusion backbone, thereby strengthening the model's audio understanding capabilities. LoRA strikes a balance between adaptation capacity and training efficiency while effectively mitigating catastrophic forgetting. The formulation of the SFT loss applied across the first three stages is detailed in Section~\ref{sec:sft}.

\paragraph{Stage 4: Preference optimization with VRPO.}
Finally, to refine instruction following and acoustic sensitivity, we apply VRPO~\citep{llada1.5} in Stage~4. Unlike standard direct preference optimization (DPO)~\citep{rafailov2023direct}, which can suffer from high variance in the evidence lower bound (ELBO) estimates for dLLMs, VRPO employs variance reduction to stabilize training (details in Sec.~\ref{sec:vrpo}). Using chosen--rejected AQA pairs, DIFFA-2 learns to prefer responses that are more faithful to subtle audio cues (e.g., prosody, emotion, background sound) from curated preference data, further enhancing its audio understanding capabilities.

\subsection{Supervised Fine-Tuning of DIFFA-2}
\label{sec:sft}

Given a textual prompt $x$, an audio input $a$, and a target response $y$, DIFFA-2 models $p_\theta(y \mid x, a)$ through a diffusion denoising process over text tokens, with audio embeddings inserted into the prompt. Audio embeddings (from the adapters) and text prompt tokens remain fully visible and are never masked; only the response tokens are corrupted and denoised.

Let $r_0$ denote the full response sequence, with length $L'$. During training, the special token \texttt{<endoftext>} is used both as padding and as the end-of-sequence marker, and the model is required to predict it. At each training iteration, tokens in $r_0$ are independently masked by a special mask token $\mathrm{M}$ with probability $t \in (0,1]$, yielding a noised response $r_t$. The model is then optimized using a diffusion-style masked prediction objective:
\begin{align}
\label{eq:sft-loss}
\mathcal{L}_{sft-a} =  - \mathbb{E}_{t, a, p, r_0, r_t} \nonumber
 \left[\frac{1}{t} \sum_{i=1}^{L'} \mathbf{1}[r_t^i = \textrm{M}] 
\log p_{\theta}(r_0^i \mid a, p, r_t) \right],
\end{align}
where $r_t$ represents the masked response corresponding to a masking ratio of $t$.
This loss trains DIFFA-2 to reconstruct masked tokens in $r_t$ given both audio and textual context, enabling the backbone to exploit bidirectional context and multimodal cues.

\subsection{Variance-Reduced Preference Optimization}
\label{sec:vrpo}
For preference alignment, we adopt VRPO proposed in \citep{llada1.5}. This method leverages a DPO-style objective function with Monte Carlo estimates of  ELBO, and incorporates optimal budget allocation as well as antithetic sampling strategies to explicitly reduce the variance of these estimates. As illustrated in Figure~\ref{pic:train}, we perform $N$ (=4) independent sampling steps, and crucially, share identical masking patterns between the policy model and the reference model when calculating $\widehat{\log p_\theta}$ and $\widehat{\log p_{\mathrm{ref}}}$. This design effectively implements antithetic sampling over diffusion trajectories, thereby stabilizing the preference learning process even when handling long, acoustically rich input sequences.

We first estimate log-likelihoods $\widehat{\log p_\theta}(y \mid x,a)$ using Monte Carlo ELBO estimation:
\begin{equation}
\widehat{\log p_\theta}(y \mid x,a)
=
\frac{1}{K}\sum_{k=1}^{K}\mathrm{ELBO}^{(k)}_\theta(y \mid x,a),
\end{equation}
where K is the sample budget.
We analogously obtain $\widehat{\log p_{\mathrm{ref}}}(y \mid x,a)$ using a frozen reference model.
VRPO then applies a DPO-style objective to the estimated log-ratios:
\begin{align}
s_\theta(y) &= \widehat{\log p_\theta}(y \mid x,a) - \widehat{\log p_{\mathrm{ref}}}(y \mid x,a), \\
\mathcal{L}_{\mathrm{VRPO}}
&=
-\log \sigma\!\Big(\beta\big[s_\theta(y^+) - s_\theta(y^-)\big]\Big),
\end{align}
where $y^+$ and $y^-$ denote the preferred and rejected responses, respectively, and $\beta$ controls the strength of preference enforcement.

\subsection{Inference Procedure}

At inference time, we pad the prompt and audio input to the target length and initialize the response sequence $r_T$ as fully masked. DIFFA-2 then performs iterative denoising over $T$ steps, gradually refining $r_t$ from coarse to fine. At each transition $t \rightarrow s$, the model predicts the masked tokens conditioned on the audio input $a$, prompt $p$, and current corrupted sequence $r_t$:
\begin{equation}
\hat{r}_t = \arg\max p_\theta(r_0 \mid a, p, r_t),
\end{equation}
and re-masks the lowest-confidence fraction of tokens to form $r_s$, enabling iterative refinement with bidirectional context. Following LLaDA~\citep{llada}, we adopt a semi-autoregressive strategy that decodes the response in left-to-right blocks, while predicting tokens within each block in parallel and selectively re-masking them across diffusion steps.

To further accelerate decoding, we use the factor-based parallel decoding strategy from fast-dLLMs~\citep{fastdllm}, which adaptively chooses how many tokens to update in parallel based on model confidence rather than a fixed threshold. Intuitively, the algorithm allows more aggressive parallel decoding when the model is confident and reduces parallelism in uncertain regions. The formal decision rule and implementation details are provided in Appendix~\ref{appendix:inference}.

\begin{table*}[!t]
\centering
\footnotesize
\renewcommand{\arraystretch}{1.0}
\setlength{\tabcolsep}{4pt}
\caption{Performance breakdown on the MMSU benchmark across perception and reasoning dimensions in Semantics (Seman.), Phonology (Phono.), and Paralinguistics (Para.) domains. "w/ FPD" denotes factor-based parallel decoding.}
\label{tab:mmsu}
\resizebox{0.95\linewidth}{!}{
\begin{tabular}{lc|cccc|cccc|c}
\toprule
\multirow{2.5}{*}{\textbf{Models}}
& \multirow{2.5}{*}{\textbf{Size}}
& \multicolumn{4}{c|}{\textbf{Perception}}
& \multicolumn{4}{c|}{\textbf{Reasoning}}
& \multirow{2.5}{*}{\textbf{Overall}}
\\
\cmidrule(lr){3-6} \cmidrule(lr){7-10}
&
& \textbf{Seman.} & \textbf{Phono.} & \textbf{Para.} & \textbf{Avg}
& \textbf{Seman.} & \textbf{Phono.} & \textbf{Para.} & \textbf{Avg}
&
\\
\midrule
\textcolor{gray}{Qwen3-Omni} & \textcolor{gray}{30B-A3B} & \textcolor{gray}{72.13} & \textcolor{gray}{55.83} & \textcolor{gray}{38.85} & \textcolor{gray}{53.20} & \textcolor{gray}{86.64} & \textcolor{gray}{82.19} & \textcolor{gray}{43.58} & \textcolor{gray}{78.88} & \textcolor{gray}{65.63} \\
\textcolor{gray}{GPT-4o-Audio}       & \textcolor{gray}{-}       & \textcolor{gray}{59.70} & \textcolor{gray}{41.56} & \textcolor{gray}{21.44} & \textcolor{gray}{39.67} & \textcolor{gray}{80.83} & \textcolor{gray}{78.74} & \textcolor{gray}{26.25} & \textcolor{gray}{71.96} & \textcolor{gray}{56.38} \\
\textcolor{gray}{Gemini   2.0 Flash} & \textcolor{gray}{-}       & \textcolor{gray}{47.17} & \textcolor{gray}{41.30} & \textcolor{gray}{30.62} & \textcolor{gray}{40.83} & \textcolor{gray}{70.69} & \textcolor{gray}{70.69} & \textcolor{gray}{36.16} & \textcolor{gray}{47.83} & \textcolor{gray}{51.03} \\
\midrule
\rowcolor{gray!30} DIFFA-2            & 8B      & \underline{60.63} & 39.04 & \textbf{41.92} & \textbf{45.58} & \textbf{85.29} & \textbf{77.58} & 43.58 & \textbf{76.40} & \textbf{60.45} \\
DIFFA-2 (w/ FPD) & 8B & 59.69 & 39.57 & \underline{40.93} & \underline{45.06} & \underline{85.02} & \underline{77.28} & 42.09 & 76.00 & \underline{60.10} \\
Kimi-Audio         & 7B      & 57.64 & \underline{42.30} & 35.74 & 43.52 & 81.77 & 76.65 & \textbf{55.22} & \underline{76.03} & 59.28 \\
Qwen2.5-Omni      & 7B      & \textbf{61.11} & \textbf{43.96} & 33.20 & 43.97 & 82.40 & 76.77 & \underline{46.87} & 75.21 & 59.09 \\
MiniCPM-O          & 8B      & 56.56 & 34.05 & 36.48 & 40.54 & 80.71 & 74.72 & 46.71 & 73.57 & 56.53 \\
\rowcolor{gray!10} DIFFA              & 8B      & 52.67 & 36.65 & 35.12 & 40.28 & 81.53 & 72.68 & 45.67 & 72.92 & 56.04 \\
Qwen2-Audio        & 8B      & 52.14 & 32.87 & 35.56 & 39.02 & 77.62 & 64.81 & 46.67 & 68.90 & 53.27 \\
Qwen-Audio-Chat    & 8B      & 57.21 & 38.52 & 24.70 & 35.69 & 58.61 & 59.78 & 25.60 & 55.93 & 46.92 \\
Phi-4-multimodal   & 8B      & 38.72 & 34.86 & 29.56 & 33.41 & 57.81 & 65.94 & 42.09 & 57.59 & 44.96 \\
Baichuan-Audio     & 11B     & 39.63 & 31.26 & 27.09 & 31.48 & 57.96 & 63.92 & 34.35 & 55.70 & 43.09 \\
GLM-4-Voice        & 9B      & 27.80 & 24.52 & 27.34 & 26.18 & 46.10 & 48.16 & 44.35 & 46.76 & 35.51 \\
Salmonn            & 7B      & 31.55 & 29.08 & 28.71 & 29.83 & 36.43 & 26.22 & 25.26 & 30.04 & 30.01 \\
LTU                & 7B      & 21.34 & 22.46 & 18.73 & 20.81 & 22.65 & 25.53 & 24.74 & 24.37 & 22.61 \\

\bottomrule
\end{tabular}
}

\end{table*}

\begin{table*}[!t]
\centering
\small
\caption{Performance breakdown on the MMAU benchmark. "w/ FPD" denotes factor-based parallel decoding.}
\label{tab:mmau}
\renewcommand{\arraystretch}{1.00}
\resizebox{0.95\linewidth}{!}{
\begin{tabular}{lccccccccc}
\toprule
\multirow{2.5}{*}{\textbf{Model}} 
& \multirow{2.5}{*}{\textbf{Size}} 
& \multicolumn{2}{c}{\textbf{Sound}} 
& \multicolumn{2}{c}{\textbf{Music}} 
& \multicolumn{2}{c}{\textbf{Speech}} 
& \multicolumn{2}{c}{\textbf{Avg}} 
\\
\cmidrule(lr){3-4} \cmidrule(lr){5-6} \cmidrule(lr){7-8} \cmidrule(lr){9-10}
& 
& \textbf{Test-mini} & \textbf{Test} 
& \textbf{Test-mini} & \textbf{Test} 
& \textbf{Test-mini} & \textbf{Test} 
& \textbf{Test-mini} & \textbf{Test}
\\ 
\midrule
\textcolor{gray}{Qwen3-Omni} & \textcolor{gray}{30B-A3B} & \textcolor{gray}{78.68} & \textcolor{gray}{73.70} & \textcolor{gray}{69.46} & \textcolor{gray}{72.22} & \textcolor{gray}{69.37} & \textcolor{gray}{66.46} & \textcolor{gray}{72.50} & \textcolor{gray}{70.77} \\
\textcolor{gray}{Gemini 2.0 Flash}       & \textcolor{gray}{-}                     & \textcolor{gray}{71.17}         & \textcolor{gray}{68.93}     & \textcolor{gray}{65.27}         & \textcolor{gray}{59.30}     & \textcolor{gray}{75.08}          & \textcolor{gray}{72.87}     & \textcolor{gray}{70.50}        & \textcolor{gray}{67.03}    \\
\textcolor{gray}{GPT-4o-Audio}           & \textcolor{gray}{-}                     & \textcolor{gray}{64.56}         & \textcolor{gray}{63.20}     & \textcolor{gray}{56.29}         & \textcolor{gray}{49.93}     & \textcolor{gray}{66.67}          & \textcolor{gray}{69.33}     & \textcolor{gray}{62.50}        & \textcolor{gray}{60.82}    \\
\midrule
\rowcolor{gray!30}
DIFFA-2                 & 8B                    &\textbf{ 76.28}         & \textbf{70.83}     & 63.47         & 60.10     & \textbf{69.06}          & \textbf{70.18}     & \textbf{69.60}        & \textbf{67.00}    \\
Qwen2.5-Omni         & 7B                    & 72.97         & 69.53     & 61.68         & \underline{62.50}     & 60.96          & 67.93     & 65.20        & \underline{66.64}    \\
DIFFA-2 (w/ FPD)   & 8B & 75.38 & 69.43 &61.08 & 59.57 & \underline{68.47} & \underline{70.15} & \underline{68.30} & 66.34 \\
Kimi-Audio             & 7B                    & \underline{75.68}         & \underline{70.70}     & \textbf{66.77}        & \textbf{65.93}     & 62.16          & 56.57     & 68.20       & 64.40    \\
MiniCPM-O               & 8B                    & 71.47        & -          & \underline{65.57}             & -         & 63.06             & -         & 66.70     & -   \\
Phi-4-multimodal       & 8B                    & 65.47         & 62.67     & 64.37         & 61.97     & 67.27          & 63.80     & 65.70        & 62.81    \\
Qwen2-Audio            & 8B                    & 67.27         & 61.17     & 56.29         & 55.67     & 55.26          & 55.37     & 59.60        & 57.40    \\
\rowcolor{gray!10} DIFFA                  & 8B                    & 46.25         & -         & 43.41         & -         & 59.46          & -         & 49.71        & -        \\
Baichuan-Audio         & 11B                   & -             & 59.46     & -             & 49.10     & -              & 42.47     & -            & 50.34    \\
Qwen-Audio-Chat        & 8B                    & 55.25         & 56.73     & 44.00         & 40.90     & 30.03          & 27.95     & 43.10        & 41.86    \\
Salmonn                & 7B                    & 41.14         & 42.10     & 37.13         & 37.83     & 26.43          & 28.77     & 34.90        & 36.23    \\
GLM-4-Voice            & 9B                    & -             & 27.63     & -             & 27.84     & -              & 35.44     & -            & 30.30    \\
LTU                    & 7B                    & 20.42         & 20.67     & 15.97         & 15.68     & 15.92          & 15.33     & 17.44        & 17.23   \\
\bottomrule
\end{tabular}
}
\end{table*}
\section{Experimental Setup}

\subsection{Training and Inference Setup}

DIFFA-2 is trained with the four-stage curriculum in Section~\ref{sec:method} using only fully open-source corpora. In total, we use about 11{,}000 hours of ASR data in Stage~1 and 3{,}767 hours of curated supervised fine-tuning data in Stages~2--3, together with roughly 3{,}000 preference pairs for Stage~4; a detailed breakdown of datasets, sample counts, and prompts is given in Appendix~\ref{appendix:data-details}. We adopt LLaDA-8B-Instruct as the dLLM backbone and update only lightweight components (semantic adapter, acoustic adapter, and LoRA parameters), resulting in roughly \textbf{1.1\%} trainable parameters overall (see Table~\ref{tab:trainable_param}). Inference configurations for each benchmark are summarized in Appendix~\ref{appendix:inference}.

\subsection{Baselines}

We compare DIFFA-2 with both proprietary and open-source audio LLMs. 
As proprietary reference points, we include GPT-4o-Audio~\citep{Gpt-4o-card} and Gemini 2.0 Flash~\citep{gemini}. 
Among open-source models, we focus on strong omni/audio baselines such as Qwen3-Omni~\citep{Qwen3-Omni}, Qwen2.5-Omni~\citep{Qwen2.5-0mni}, Kimi-Audio~\citep{Kimi-audio}, and the first-generation DIFFA~\citep{diffa}. 
The full list of baselines is given in Appendix~\ref{appendix:baselines}.

\subsection{Benchmarks}

We evaluate DIFFA-2 on four representative benchmarks. Among them, \textbf{MMSU}~\citep{wang2025mmsu}, \textbf{MMAU}~\citep{sakshi2025mmau}, and \textbf{MMAR}~\citep{ma2025mmar} are audio understanding benchmarks and constitute the primary focus of our study, while \textbf{VoiceBench}~\citep{chen2024voicebench} is included only as an auxiliary evaluation of semantic dialogue ability. Further details on benchmarks are provided in Appendix~\ref{appendix:benchmark}.

\section{Experiments}

\subsection{Results on Benchmarks}

\paragraph{MMSU.}
Table~\ref{tab:mmsu} reports a fine-grained breakdown on MMSU.
Among open LALMs of comparable size, DIFFA-2 attains the best overall accuracy (60.45), slightly outperforming Kimi-Audio (59.28), Qwen2.5-Omni (59.09), and other 7--11B baselines, while remaining within roughly 5 points of the larger proprietary Qwen3-Omni.
DIFFA-2 also achieves the highest perception average (45.58), with clear gains on paralinguistic perception (41.92) and solid performance on semantic and phonological perception.
On reasoning, DIFFA-2 reaches the best average (76.40) among open models, with noticeable improvements in semantic and phonological reasoning compared with Kimi-Audio and Qwen2.5-Omni.
Relative to the first-generation DIFFA, DIFFA-2 improves both perception and reasoning (overall +4.41 points), indicating that the upgraded acoustic front-end and four-stage training pipeline translate into stronger speech understanding on MMSU.

\begin{table*}[!t]
\centering
\tiny
\caption{Performance breakdown on the MMAR benchmark. The 'All' category denotes the comprehensive evaluation across mixed sound, music, and speech modalities. "w/ FPD" denotes factor-based parallel decoding.}
\label{tab:mmar}
\resizebox{\linewidth}{!}{
\begin{tabular}{lccccccccc}
\toprule 
\multirow{2.5}{*}{\textbf{Models}} & \multirow{2.5}{*}{\textbf{Size}} & \multicolumn{3}{c}{\textbf{Single Modality (\%)}} &  \multicolumn{4}{c}{\textbf{Mixed Modalities (\%)}} & \multirow{2.5}{*}{\textbf{Avg (\%)}} \\ 
\cmidrule(lr){3-5} \cmidrule(lr){6-9}
&& \textbf{Sound} & \textbf{Music} & \textbf{Speech} & \textbf{Sound-Music} & \textbf{Sound-Speech} & \textbf{Music-Speech} & \textbf{All} & \\
\midrule
\textcolor{gray}{Qwen3-Omni} & \textcolor{gray}{30B-A3B} & \textcolor{gray}{59.39} & \textcolor{gray}{54.37} & \textcolor{gray}{70.41} & \textcolor{gray}{90.91} & \textcolor{gray}{74.77} & \textcolor{gray}{63.41} & \textcolor{gray}{70.83} & \textcolor{gray}{65.90} \\
\textcolor{gray}{Gemini 2.0 Flash}  & \textcolor{gray}{-}          & \textcolor{gray}{61.21} & \textcolor{gray}{50.97} & \textcolor{gray}{72.11}  & \textcolor{gray}{81.82}       & \textcolor{gray}{72.48}        & \textcolor{gray}{65.85}        & \textcolor{gray}{70.83}              & \textcolor{gray}{65.60} \\
\textcolor{gray}{GPT-4o-Audio}      & \textcolor{gray}{-}          & \textcolor{gray}{53.94} & \textcolor{gray}{50.97} & \textcolor{gray}{70.41}  & \textcolor{gray}{63.64}       & \textcolor{gray}{72.48}        & \textcolor{gray}{62.20}        & \textcolor{gray}{75.00}              & \textcolor{gray}{63.50} \\

\midrule
Qwen2.5-Omni & 7B & \textbf{55.76} & \textbf{41.75} & \textbf{54.42} & \textbf{45.45} & 55.96 & \textbf{57.32} & \textbf{54.17} & \textbf{51.40} \\
\rowcolor{gray!30} DIFFA-2 & 8B & \underline{54.55} & \textbf{41.75} & \underline{53.40} & \textbf{45.45} & \textbf{58.26} & \underline{54.88} & 37.50 & \underline{50.80} \\
DIFFA-2 (w/ FPD) &8B & 53.94 & 36.41 & 52.72 & 36.36 & \underline{56.88} & 53.66 & 45.83 & 50.20 \\
MiniCPM-O & 8B & 49.70 & 36.41 & 50.00 & 36.36 & 53.67 & 45.12 & \textbf{54.17} & 48.60 \\
Baichuan-Omni-1.5 & 7B & 41.21 & 33.01 & 40.48 & 36.36 & 48.62 & 39.02 & 41.67 & 40.70 \\
\rowcolor{gray!10} DIFFA & 8B & 37.58 & 31.07 & 39.46 & 36.36 & 43.12 & 45.12 & 25.00 & 37.20 \\
Salmonn & 13B & 30.30 & 29.61 & 34.69 & 9.09 & 34.86 & 35.37 & 41.67 & 33.20 \\
Salmonn & 7B & 30.91 & 25.73 & 34.35 & 9.09 & 37.61 & 28.05 & 37.50 & 32.80 \\
Qwen2-Audio & 8B & 33.33 & 24.27 & 32.31 & 9.09 & 31.19 & 30.49 & 25.00 & 30.00 \\
OpenOmni & 8B & 20.61 & 22.33 & 35.37 & 18.18 & 27.06 & 23.17 & 25.00 & 27.00 \\
Qwen-Audio-Chat & 8B & 27.88 & 20.39 & 22.11 & 9.09 & 25.23 & 25.61 & 20.83 & 23.50 \\
LTU & 7B & 19.39 & 19.90 & 13.95 & 18.18 & 24.77 & 21.95 & 16.67 & 19.20 \\
\bottomrule
\end{tabular}}
\end{table*}

\begin{table*}[!t]
\centering
\caption{Word error rate (WER\%) and real-time factor (RTF) of Stage-1 with dLLMs and AR backbone on Librispeech-clean and Librispeech-other testing set. "w/ FPD" denotes factor-based parallel decoding.}
\label{tab:asr-ablation}
\resizebox{0.5\linewidth}{!}{
\begin{tabular}{lcccc}
\toprule
\multirow{2}{*}{Model}          & \multicolumn{2}{c}{WER $\downarrow$	} & \multicolumn{2}{c}{RTF $\downarrow$	} \\
\cmidrule(lr){2-3} \cmidrule(lr){4-5} 
                                & clean      & other      & clean      & other      \\
                                \midrule
LLaMA-Audio (S1)                 & \textbf{2.43}       & \textbf{5.09}       & 0.1402 & 0.1418 \\
DIFFA-2 (S1)                      & 2.72       & 5.34       & 0.6792  & 0.7489          \\
DIFFA-2 (S1 w/ FPD) & 3.05       & 5.68       & \textbf{0.0820} & \textbf{0.0867} \\
\bottomrule
\end{tabular}
}

\end{table*}

\paragraph{MMAU.}
On MMAU (Table~\ref{tab:mmau}), DIFFA-2 again performs competitively with strong AR LALMs.
It achieves the best average accuracy among open models on both \textit{Test-mini} and \textit{Test} splits (69.60 and 67.00) surpassing Qwen2.5-Omni and Kimi-Audio, and approaching larger or proprietary systems such as Qwen3-Omni and Gemini.
DIFFA-2 is particularly strong on sound and speech, where it attains the highest scores among open models, while its music performance is slightly behind Kimi-Audio and MiniCPM-O but remains competitive without any music-specialized design.
Compared with DIFFA, DIFFA-2 gains nearly 20 points in average accuracy on \textit{Test-mini} (49.71 $\rightarrow$ 69.60), suggesting that the enhanced acoustic modeling and multi-stage training generalize well to high-level audio understanding across sound, music, and speech.

\paragraph{MMAR.}
The MMAR benchmark (Table~\ref{tab:mmar}) evaluates single and mixed audio modalities with more compositional queries.
DIFFA-2 achieves an average accuracy of 50.80\%, substantially improving over DIFFA (37.20\%, +13.6 points) and outperforming other 8B open baselines such as MiniCPM-O (48.60\%) and Qwen2-Audio (30.00\%).
On single-modality tasks (sound, music, speech), DIFFA-2 consistently improves over DIFFA and narrows the gap to Qwen2.5-Omni.
For the most challenging Sound--Music--Speech mixtures, DIFFA-2 underperforms relative to Qwen2.5-Omni, likely reflecting the lack of mixed-modality supervision in the training data.
Overall, MMAR indicates that DIFFA-2 effectively extends diffusion-based modeling to multi-source audio composition, while complex three-way mixtures remain a challenging regime compared with the strongest AR-based LALMs.

\begin{table*}[!t]
\centering
\caption{Ablation of DIFFA-2 and LLaMA-Audio's multi-stage training on MMAU and MMSU. LLaMA-Audio is based on LLaMA 3.1 backbone and then trained with the same data and settings.}
\label{tab:stage-ablation}
\resizebox{\linewidth}{!}{
\begin{tabular}{l|cccc|ccccccccc}
\toprule
\multirow{2.5}{*}{\textbf{Model}} 
& \multicolumn{3}{c}{\textbf{MMAU}} 
& \multirow{2.5}{*}{\textbf{ Overall}}
& \multicolumn{4}{c}{\textbf{MMSU (Perception)}} 
& \multicolumn{4}{c}{\textbf{MMSU (Reasoning)}} 
& \multirow{2.5}{*}{\textbf{ Overall}} \\
\cmidrule(lr){2-4} \cmidrule(lr){6-9} \cmidrule(lr){10-13}
& \textbf{Sound} & \textbf{Music} & \textbf{Speech}
& 
& \textbf{Sem.} & \textbf{Phon.} & \textbf{Para.} & \textbf{Avg}
& \textbf{Sem.} & \textbf{Phon.} & \textbf{Para.} & \textbf{Avg}
& \\
\midrule
LLaMA-Audio (S2) & 65.47 & 54.79 & 62.16 & 60.80 & 39.21 & 30.91 & 30.23 & 32.69 & 51.35 & 63.97 & 44.18 & 55.45 & 43.71 \\
LLaMA-Audio (S3) & 73.87 & 65.87 & 62.46 & 67.40 & 50.39 & \textbf{39.89} & 36.67 & 41.22 & 75.45 & 73.18 & 45.07 & 70.33 & 55.31 \\
\midrule
DIFFA-2 (S2)     & 72.07 & 52.69 & 66.97 & 63.90
                      & 52.91 & 36.58 & 31.32 & 38.54
                      & 84.03 & 75.74 & \textbf{46.57} & 75.50
                      & 56.43 \\
DIFFA-2 (S3)     & 74.77 & 62.57 & 67.27 & 68.20
                      & 59.53 & 37.54 & 40.63 & 44.16
                      & 84.66 & 76.15 & 44.48 & 75.70
                      & 59.41 \\
DIFFA-2 (S4)     & \textbf{76.28} & \textbf{63.47} & \textbf{69.07} &\textbf{ 69.60}
                      & \textbf{60.63} & 39.04 & \textbf{41.92} & \textbf{45.58}
                      & \textbf{85.29} & \textbf{77.58} & 43.58 & \textbf{76.40}
                      & \textbf{60.45} \\
\bottomrule
\end{tabular}
}

\end{table*}

\paragraph{VoiceBench.}
We additionally evaluate on VoiceBench to assess dialogue-style spoken interaction.
As shown in Appendix~\ref{appendix:addition_exp}, DIFFA-2 lags behind heavily instruction-tuned omni models such as GPT-4o-Audio and Qwen3-Omni, but still improves over DIFFA and several open-source baselines, which is consistent with our design focus on audio understanding rather than extensive conversational tuning.

\paragraph{Summary.}
Across MMSU, MMAU, and MMAR, DIFFA-2 consistently outperforms the first-generation DIFFA and often matches or surpasses strong open AR LALMs of similar size, while approaching larger proprietary systems on several metrics. 
Gains are particularly clear in semantic and phonological reasoning and in sound and speech understanding, whereas performance on VoiceBench remains behind omni models that are heavily tuned for dialogue and alignment.
These results indicate that, under realistic data constraints, diffusion-based backbones can serve as competitive audio understanding models, and that additional dialogue-centric supervision could further close the gap on interactive voice assistant benchmarks.

\subsection{Ablation Study}

We first compare Stage-1 ASR performance between diffusion and autoregressive backbones (Table~\ref{tab:asr-ablation}). Under the same ASR-style training on LibriSpeech, the AR baseline LLaMA-Audio (S1) attains slightly lower word error rate (WER) than DIFFA-2 (S1), which is consistent with the advantage of strictly left-to-right decoding for monotonic transcription. When we enable factor-based parallel decoding for DIFFA-2 (S1, w/ FPD), WER increases moderately, but the real-time factor (RTF) drops substantially and becomes lower than that of the AR baseline. This indicates that a diffusion backbone does not inherently improve low-level recognition accuracy, but its inference latency can be made competitive with, or better than, an AR backbone by using an appropriate parallel decoding scheme in ASR task. 
On audio understanding benchmarks, DIFFA-2 with factor-based parallel decoding attains accuracy that is close to the standard setting, suggesting that it provides a practical knob to trade off accuracy and latency for diffusion-based audio models.

Table~\ref{tab:stage-ablation} reports the multi-stage training ablation on MMAU and MMSU, comparing DIFFA-2 with LLaMA-Audio under matched data and training curriculum. With only Stage~2 (adapter-only alignment), DIFFA-2 already achieves higher overall scores than LLaMA-Audio (S2) on both benchmarks, suggesting that the diffusion backbone benefits more from the same semantic--acoustic alignment. Advancing from Stage~2 to Stage~3 improves both models, and DIFFA-2 gains more on reasoning-oriented metrics. Adding Stage~4 (VRPO) further improves DIFFA-2, yielding the best overall performance and more balanced gains across perception and reasoning.

Overall, the ablation highlights a difference between transcription and audio understanding. For token-level ASR, the AR backbone retains a small advantage in WER, in line with its sequential decoding nature. For holistic audio QA, DIFFA-2 with a diffusion backbone achieves stronger performance under the same data and multi-stage training, which may be attributed to the corruption–reconstruction training objective of dLLMs, which has been shown in text domains to make more effective use of limited data, making it better aligned with audio understanding benchmarks, although we do not completely disentangle backbone pre-training effects.

\section{Related Work}

\subsection{Large Audio Language Models}

Recent LALMs primarily adopt autoregressive backbones. A common design couples a speech encoder to an LLM via lightweight bridging modules (e.g., Qwen2-Audio~\citep{Qwen2-Audio}, SALMONN~\citep{tang2024salmonn}, Audio-Flamingo2~\citep{Audio-Flamingo-2}, with omni models further supporting streaming and multi-modality~\citep{Qwen2.5-0mni, Qwen3-Omni}. Another line tokenizes audio into discrete sequences~\citep{zhang2023speechgpt,defossez2024moshi}, while Kimi-Audio~\citep{Kimi-audio} fuses discrete and continuous representations. These models largely rely on AR decoding.

\subsection{Diffusion Large Language Models}

Diffusion large language models generate sequences by iteratively denoising corrupted tokens, offering bidirectional context modeling and parallel token updates. Early work on diffusion for discrete text~\citep{Austin_Johnson_Ho_Tarlow_Berg_2021,shi2024simplified,Sahoo2024SimpleAE} established the feasibility of this paradigm, and LLaDA~\citep{llada,llada1.5} scaled it to large language models with strong performance on understanding and reasoning tasks. Recent efforts improve inference efficiency via training-free acceleration, including KV-cache-like reuse with confidence-aware parallel decoding and adaptive length prediction~\citep{li2025beyond,odb-dllm}. Our work is closely related to diffusion LMs such as LLaDA and fast-dLLMs~\citep{fastdllm}. LLaDA and its variants focus on text generation and do not consider audio encoders or audio-specific training curricula.

\section{Conclusions}

This paper presents \textbf{DIFFA-2}, an enhanced dLLMs-based LALM for audio understanding. Despite having only 1.1\% (99M) trainable parameters and utilizing a modest 14.8k hours of open-source data, DIFFA-2 achieves substantial performance gains over its predecessor. Evaluations on MMSU, MMAU, and MMAR benchmarks demonstrate that DIFFA-2 is competitive with leading autoregressive models. These results establish dLLMs-based modeling as a highly competitive alternative for universal audio understanding tasks.

\section*{Limitations}
\label{sec:limitations}
Although DIFFA-2 achieves strong results on audio understanding benchmarks, several limitations remain. 
First, our training objectives and data curation are geared toward fine-grained audio understanding rather than open-domain spoken dialogue. 
Consequently, DIFFA-2 is exposed to only limited conversational and alignment-style supervision, which is reflected in its mid-range performance on VoiceBench compared with heavily instruction-tuned AR Omni-models. 
Designing a more balanced training recipe that jointly targets audio understanding and spoken dialogue is an important direction for future work.
Second, we focus exclusively on text-based audio understanding and do not consider speech generation or streaming, full-duplex interaction. 
DIFFA-2 is evaluated in an offline speech-in/text-out setting; integrating it into end-to-end speech-in/speech-out systems and assessing user-centric metrics such as latency and interaction quality are important next steps.
Third, we apply a simple training-free factor-based parallel decoding scheme to reduce diffusion steps and observe clear latency gains with negligible accuracy loss, but DIFFA-2 is not yet uniformly faster than strong AR audio LLMs under all settings. 
We view this as a systems-level design choice rather than a fundamental limitation of dLLMs backbones; adapting more advanced training-free acceleration methods from text dLLMs to audio is a promising but orthogonal direction for future work.

\bibliography{iclr2026_conference}
\bibliographystyle{iclr2026_conference}

\newpage
\appendix

\setcounter{figure}{0}  
\setcounter{table}{0}   
\renewcommand{\thefigure}{A.\arabic{figure}}
\renewcommand{\thetable}{A.\arabic{table}}

\section{Data Details and Prompt Templates}
\label{appendix:data-details}

\subsection{ASR Data}
We use LibriSpeech~\citep{panayotov2015librispeech} and GigaSpeech~\citep{chen2021gigaspeech} as ASR corpora. For each transcript, we construct instruction-style ASR samples by applying 25 instruction templates generated by Qwen-32B (e.g., ``Please transcribe the following audio...'' or ``What is the exact content of this recording?''). The full list of instruction templates is shown in Table~\ref{fig:stage1_asr_prompt}.

\begin{figure}[ht]
    \centering
    \begin{promptbox}[Examples of ASR Prompts in Stage 1]
    \begin{lstlisting}[style=promptstyle]
1. Please transcribe the audio to text.
2. Convert this speech to text.
3. What is being said in this audio?
4. Transcribe the following audio clip.
5. Please write down what you hear in the audio.
6. Convert the spoken words to written text.
7. What words are spoken in this recording?
8. Please provide a transcription of this audio.
9. Turn this speech into text format.
10. Write out what is said in the audio file.
    \end{lstlisting}
    \end{promptbox}
    \caption{Examples of ASR Prompts in Stage 1}
    \label{fig:stage1_asr_prompt}
\end{figure}
\subsection{SFT Data}

\begin{table}[!t]
  \centering
  \small
  \caption{Dataset Statistics categorized by domain (Sound, Music, and Speech). Duration represents unique audio hours.}
  \label{tab:dataset_stats_grouped}

  \resizebox{0.6\linewidth}{!}{
  \begin{tabular}{lrr}
    \toprule
    \textbf{Dataset} & \textbf{Samples} & \textbf{Duration (h)} \\
    \midrule
    \multicolumn{3}{l}{\textit{\textbf{Sound \& General Audio}}} \\
    AudioCaps & 181,453 & 383.47 \\
    Clotho & 75,090 & 18.04 \\
    ESC50 & 9,000 & 1.39 \\
    TACOS & 33,320 & 57.78 \\
    VocalSound & 20,208 & 23.51 \\
    WavCaps\_AudioSetSL & 108,308 & 296.95 \\
    WavCaps\_Freesound30s & 155,287 & 432.80 \\
    \textbf{\textit{Subtotal}} & \textbf{\textit{582,666}} & \textbf{\textit{1,213.94}} \\
    \midrule
    \multicolumn{3}{l}{\textit{\textbf{Music}}} \\
    FMA\_medium & 16,896 & 140.74 \\
    LP-MusicCaps-MTT & 15,560 & 126.54 \\
    MusicCaps & 2,568 & 7.14 \\
    Nsynth & 296,382 & 329.31 \\
    \textbf{\textit{Subtotal}} & \textbf{\textit{331,406}} & \textbf{\textit{603.73}} \\
    \midrule
    \multicolumn{3}{l}{\textit{\textbf{Speech}}} \\
    AccentDB\_extended & 50,622 & 19.28 \\
    CompA-R & 197,218 & 170.60 \\
    EmoV\_DB & 68,930 & 9.49 \\
    IEMOCAP & 50,304 & 5.23 \\
    MELD & 6,450 & 0.93 \\
    SpeechCraft & 228,008 & 483.82 \\
    VCTK-Corpus & 176,968 & 44.04 \\
    AlpacaTrain & 19,307 & 25.17 \\
    MetaFairASR & 46,342 & 48.66 \\
    NaturalQuestions & 9,022 & 10.25 \\
    Opens2s & 49,368 & 67.47 \\
    Paraspeechcaps & 184,552 & 497.07 \\
    TrivalQA & 79,565 & 104.29 \\
    VoxCeleb1 & 297,284 & 340.39 \\
    WebQuestions & 2,576 & 2.49 \\
    \textbf{\textit{Subtotal}} & \textbf{\textit{1,466,516}} & \textbf{\textit{1,829.18}} \\
    \midrule
    \textbf{Total} & \textbf{2,380,588} & \textbf{3,646.85} \\
    \bottomrule
  \end{tabular}
  }
\end{table}

\paragraph{Part 1: Audio-caption-based AQA.}
We collect a diverse set of audio and speech datasets covering speech, environmental sounds, and music, including ParaSpeechCaps~\citep{paraspeechcaps}, AudioCaps~\citep{audiocaps}, WavCaps~\citep{wavcaps}, VocalSound~\citep{vocalsound}, NSynth~\citep{nsynth}, FMA-medium~\citep{fma_medium}, ESC-50~\citep{esc50}, Clotho~\citep{clotho}, AccentDB~\citep{accentdb}, EmoV-DB~\citep{emov_db}, IEMOCAP~\citep{iemocap}, MELD~\citep{meld}, VCTK~\citep{vctk}, Meta FAIR ASR~\citep{meta_fair_asr}, and VoxCeleb1~\citep{voxceleb1}.  
From the Desta~2.5~\citep{lu2025desta25audiogeneralpurposelargeaudio} dataset, we select a subset of question types and paralinguistic annotations, and use Qwen3-32B to generate an answer for each audio–question pair. The exact prompts are listed in Figure~\ref{fig:qa_creation_prompt}.

\begin{figure}[!t]
    \centering
\begin{promptbox}[Prompts of Audio QA Data Creation in Stage 2]
\label{prompt:aqa}
\begin{lstlisting}[style=promptstyle]
[System]
You are a helpful voice assistant.
Imagine you can hear the audio clips.
Focus on the audios and respond directly to the prompts.

[User]
This is the audio: {Audio Description}.
{Text Prompt}

[Assistant]

- ...
\end{lstlisting}
\end{promptbox}
    \caption{Prompts of Audio QA Data Creation in Stage 2}
    \label{fig:qa_creation_prompt}
\end{figure}

\paragraph{Part 2: Direct Audio QA.}
We construct three categories of direct audio QA data: simple QA, complex QA, and empathetic QA. We first collect text-only QA pairs from Alpaca~\citep{alpaca}, Natural Questions~\citep{kwiatkowski2019natural}, TriviaQA~\citep{joshi2017triviaqa}, and WebQuestions~\citep{berant2013semantic}, and then synthesize speech with CosyVoice2~\citep{du2024cosyvoice}, using speaker prompts randomly sampled from the LibriSpeech training set. We categorize samples into simple or complex based on answer length and apply different instruction templates for each type. We further augment this with the English subset of OpenS2S~\citep{wang2025opens2s}. Detailed statistics and prompt templates are shown in Tables~\ref{tab:dataset_stats_grouped} and Figure~\ref{fig:stage2_prompt}.

\begin{figure}[!t]
    \centering
\begin{promptbox}[Examples of Prompts in Stage 2]
\begin{lstlisting}[style=promptstyle]
[Simple QA Prompts]
1. Listen to the audio question and reply with a concise and factual answer.
2. Respond to the spoken query with a brief, accurate answer based on the audio.
3. Give a straightforward response to the question in the audio.
4. Respond directly to the audio question, stating only what is necessary.

[Complex QA Prompts]
1. Listen to the audio question and provide a comprehensive, detailed answer covering all aspects.
2. Respond to the spoken message with an in-depth explanation addressing every part of the query.
3. Offer a full, structured answer to the voice question, including background and supporting details.

[Sympathetic QA Prompts]
1. Listen to the voice message and respond in a natural, conversational manner, showing empathy and genuine understanding.
2. Reply to the spoken message as if engaged in a friendly, real-life conversation, maintaining warmth and authenticity.
3. Respond in a smooth, natural tone that conveys a human touch and kindness.

\end{lstlisting}
\end{promptbox}
    \caption{Examples of prompts in Stage 2}
    \label{fig:stage2_prompt}
\end{figure}

\paragraph{Part 3: ASR Data.}
We randomly sample 5\% of the ASR data used in Stage~1 and add it to the Stage~2 training set to preserve ASR-related acoustic grounding.

\paragraph{Part 4: Multi-choice AQA Data}
We employ the AudioMCQ~\citep{he2025measuring} corpus, which integrates multi-choice AQA instances derived from AudioCaps, Clotho, CompA-R~\citep{ghosh2024gama}, MusicCaps~\citep{agostinelli2023musiclm}, LP-MusicCaps~\citep{doh2023lp}, SpeechCraft~\citep{jin2024speechcraft}, and TACOS~\citep{primus2025tacos}. We follow the official data construction recipe of AudioMCQ without chain-of-thought, and then combine it with the Stage~2 data used for adapter training. Detailed dataset statistics are provided in Table~\ref{tab:dataset_stats_grouped}.

\subsection{Preference Data and Prompts}
To construct preference data for VRPO, we start from high-quality audio QA instances sampled from SFT data. Given an audio input, a question, and a reference answer, we prompt a language model to generate a fluent but partially incorrect answer that introduces subtle audio-related errors. The full prompt templates and examples for preference data construction are listed in Figure~\ref{fig:preference_prompt}.

\begin{figure}[!t]
    \centering
\begin{promptbox}[Preference Data Generation Prompt]
\begin{lstlisting}[style=promptstyle]
[System]
You are creating training data for an audio question answering model (Audio-QA).
You will NOT see the raw audio, only a textual description of it.

You are given a QUESTION and a REFERENCE_ANSWER which should be treated as the GOOD answer.
Your task is to generate ONE BAD_ANSWER to the same question.

Requirements for BAD_ANSWER:
- It must be fluent and look superficially reasonable.
- BUT it must be partially incorrect with respect to the audio description (e.g., wrong attribute, wrong event, missing key detail).
- The error should be about the AUDIO-related content (e.g., rhythm, sound type, emotion, gender), rather than just style or verbosity.
- DO NOT make the bad answer obviously nonsensical or completely unrelated.
- It must still directly answer the question (not "I don't know").

At the end, you must also CHECK whether the REFERENCE_ANSWER is clearly better than the BAD_ANSWER.
If they are too similar or you are not confident, mark the pair as unusable.

Output your result in STRICT JSON with the following fields only:
{
  "bad_answer": "...",
  "explanation": "...",
  "usable": true or false
}
Do NOT add any extra text outside the JSON object.

[User]
Audio Description:
{audio_desc}

Question:
{question}

Reference Answer (GOOD):
{gold_answer}

[Assistant]:
{bad_answer}
\end{lstlisting}
\end{promptbox}
    \caption{Preference Data Generation Prompt}
    \label{fig:preference_prompt}
\end{figure}

\setcounter{figure}{0}  
\setcounter{table}{0}   
\renewcommand{\thefigure}{B.\arabic{figure}}
\renewcommand{\thetable}{B.\arabic{table}}

\section{Additional Experimental Details}

All models and datasets utilized in this study are released under open-source licenses, in compliance with the terms of their respective original distributions.

\subsection{Baseline Models}
\label{appendix:baselines}

For completeness, we list here all baseline models used in our experiments.
In the main text, we highlight a subset of representative systems (e.g., Qwen2.5-Omni, Kimi-Audio, Qwen2-Audio, MiniCPM-O, and DIFFA); the full set of models covered in our evaluation is summarized in Table~\ref{tab:baseline-list}.

\begin{table}[h]
\centering
\small
\renewcommand{\arraystretch}{1.05}
\setlength{\tabcolsep}{5pt}
\caption{Baseline models used in our experiments.}
\label{tab:baseline-list}
\resizebox{0.5\linewidth}{!}{
\begin{tabular}{ll}
\toprule
\textbf{Model} & \textbf{Reference} \\
\midrule
GPT-4o-Audio          & \citep{Gpt-4o-card} \\
Gemini 2.0 Flash      & \citep{gemini}      \\
Qwen3-Omni            & \citep{Qwen3-Omni}  \\
Qwen2.5-Omni          & \citep{Qwen2.5-0mni}\\
Kimi-Audio            & \citep{Kimi-audio}  \\
MiniCPM-O             & \citep{MiniCPM-O}   \\
Qwen2-Audio           & \citep{Qwen2-Audio} \\
Baichuan-Omni-1.5     & \citep{Baichuan-Omni} \\
Baichuan-Audio        & \citep{Baichuan-Audio} \\
GLM-4-Voice           & \citep{GLM-4-Voice} \\
Step-Audio            & \citep{Step-Audio}  \\
LLaMA-Omni            & \citep{LLaMA-Omni}  \\
Slam-Omni             & \citep{Slam-Omni}   \\
Freeze-Omni           & \citep{freezeomni}  \\
Mini-Omni / Mini-Omni2& \citep{Mini-Omni,Mini-Omni2} \\
Moshi                 & \citep{defossez2024moshi} \\
DiVA                  & \citep{DIVA}        \\
VITA                  & \citep{VITA}        \\
LTU                   & \citep{LTU}         \\
Salmonn               & \citep{salmonn}     \\
Qwen-Audio-Chat       & \citep{qwen-audio}  \\
DIFFA                 & \citep{diffa}       \\
\bottomrule
\end{tabular}
}
\end{table}

\subsection{Benchmark Details}

\label{appendix:benchmark}
\textbf{MMSU}~\citep{wang2025mmsu} is a large-scale benchmark for assessing perception and reasoning in realistic spoken-language scenarios. It contains 5,000 audio–question–answer triplets across 47 tasks, covering both linguistic and paralinguistic phenomena such as phonetics, prosody, semantics, emotion, and speaker traits. Tasks are divided into perception- and reasoning-oriented categories, enabling a comprehensive evaluation of fine-grained audio understanding.

\textbf{MMAU}~\citep{sakshi2025mmau} evaluates advanced audio understanding via human-annotated multiple-choice questions over speech, music, and environmental sounds. It emphasizes high-level reasoning and expert knowledge rather than low-level perception. We report results on the \textit{Test-mini} split.

\textbf{MMAR}~\citep{ma2025mmar} is a multi-task audio reasoning benchmark designed to assess reasoning consistency and generalization across heterogeneous audio modalities. It focuses on joint perceptual grounding and reasoning over diverse audio inputs, providing a complementary evaluation of audio reasoning robustness.

\textbf{VoiceBench}~\citep{chen2024voicebench} targets semantic dialogue ability in spoken interaction. It is constructed by converting text-based benchmarks into audio queries using TTS and evaluates general knowledge, instruction following, and safety. Since VoiceBench primarily measures semantic dialogue performance rather than audio understanding, we include it only as a supplementary evaluation.

\subsection{Training Details}
DIFFA-2 is trained using a four-stage schedule with progressively reduced learning rates.
LoRA is applied to the dLLMs backbone in Stage~3 with a rank of 8 and a scaling factor $\alpha$ of 16, while preference optimization is performed in Stage~4. We employ 64 NVIDIA A100 GPUs for the first three stages and 4 A100 GPUs for the final stage. The entire training pipeline takes approximately 5 days to complete.
\begin{table}[t]
\centering
\small
\caption{Stage-wise training configuration of DIFFA-2.}
\label{tab:training}
\resizebox{0.6\linewidth}{!}{
\begin{tabular}{lcccc}
\toprule
\textbf{Stage} & \textbf{LR} & \textbf{Batch} & \textbf{Warmup} & \textbf{Epochs}   \\
\midrule
Stage 1 & $1\!\times\!10^{-4}$ & 1280 & 1000 & 12  \\
Stage 2 & $5\!\times\!10^{-5}$ & 196  & 1000 & 10  \\
Stage 3 & $5\!\times\!10^{-5}$ & 196  & 1000 & 10   \\
Stage 4 & $5\!\times\!10^{-6}$ & 4    & 200  & 1    \\
\bottomrule
\end{tabular}
}

\end{table}

\begin{table}[t]
\centering
\footnotesize
\caption{Statistics of parameters}
\label{tab:trainable_param}
\resizebox{0.6\linewidth}{!}{
\begin{tabular}{lcc}
\toprule
\textbf{Module} & \textbf{\#Params} & \textbf{Trainable} \\
\midrule
Whisper-Large-V3 Encoder & 637M & 0 \\

Semantic Adapter        & 36.4M & 36.4M \\
Acoustic Adapter        & 47.9M & 47.9M \\
dLLMs Backbone  & 8.03B & 0 \\
LoRA Modules            & 14.7M & 14.7M \\
\midrule
\textbf{Total}          & \textbf{8.77B} & \textbf{99.0M} \\
\bottomrule
\end{tabular}
}

\end{table}

\subsection{Inference Details}
\label{appendix:inference}
At inference time, we pad the prompt and audio input to the target length and initialize the response sequence $r_T$ as fully masked. DIFFA-2 then performs iterative denoising over $T$ steps, progressively refining the response from coarse to fine  (Figure~\ref{pic:train}).

At each denoising transition $s_1 \rightarrow s_2$, the model predicts masked tokens conditioned on the audio input $a$, prompt $p$, and the current corrupted sequence $r_t$:
\begin{equation}
\hat{r}_{s_1} = \arg\max p_\theta(r_0 \mid a, p, r_{s_1}).
\end{equation}
Token-level confidence scores are used to re-mask a fraction of low-confidence tokens—proportional to $\lceil s_2/s_1 \rceil$—to form the next intermediate sequence $r_{s_2}$, enabling iterative refinement with full bidirectional context.

Following LLaDA~\citep{llada}, we also adopt a semi-autoregressive decoding strategy that generates the response in a block-wise left-to-right manner. Within each block, tokens are decoded in parallel and selectively re-masked across diffusion steps, balancing generation quality and efficiency.

\paragraph{Factor-based parallel decoding.}
To further accelerate decoding, we adopt the factor-based parallel decoding strategy proposed in fast-dLLMs~\citep{fastdllm}. This strategy extends threshold-based decoding by adaptively determining how many tokens to decode in parallel based on model confidence, rather than relying on a fixed confidence threshold.

Given the marginal confidence estimates of candidate tokens within a decoding block, we first sort confidences in descending order and select the largest number of tokens $n$ such that
\begin{equation}
(n+1)\bigl(1 - c^{(n)}\bigr) < f,
\end{equation}
where $c^{(n)}$ denotes the $n$-th highest confidence and $f$ is a decoding factor hyperparameter. This criterion allows more aggressive parallel decoding when the model is confident, while conservatively reducing parallelism in uncertain regions.

By leveraging this factor-based strategy on top of the diffusion denoising process, DIFFA-2 achieves improved inference efficiency while preserving generation quality.

Inference hyperparameters used for each benchmark is presented in Table~\ref{tab:infer_hyperparams}.
Key parameters include the maximum answer length, the block length for incremental decoding, and the total denoising steps. We synchronize the denoising budget with the sequence length to achieve peak generation quality. For factor-based parallel decoding, we set $f$ to 1.0. 
Note that the steps are invalid when the factor parallel decoding is applied. 
All inference experiments are conducted on a single NVIDIA A100 GPU. The prompt for audio understanding tasks is provided in Figure~\ref{prompt:audio_understanding}.

\begin{figure}[!t]
    \centering
\begin{promptbox}[Prompt for Audio Understanding Tasks]

\begin{lstlisting}[style=promptstyle]
[System]
You are a helpful voice assistant.

[User]
This is the audio: {Audio Infomation}.

Choose the most suitable answer from options A, B, C, and D to respond the question in next line. Do not provide any additional explanations or content.

Question:{Text Question}

Options:{Text Options}

[Assistant]

- ...
\end{lstlisting}
\end{promptbox}
    \caption{Prompt for audio understanding tasks}
    \label{prompt:audio_understanding}
\end{figure}

\begin{table}[!h]
\small
\centering
\caption{Inference hyperparameters used for each benchmark}
\label{tab:infer_hyperparams}
\resizebox{0.6\linewidth}{!}{
\begin{tabular}{cccc}
\toprule
\textbf{Benchmark}           & \textbf{Answer length} & \textbf{Block length} & \textbf{Steps} \\
           \midrule
Librispeech &128 &128 &128 \\
\midrule
MMSU       & 16            & 16           & 16     \\
MMAU       & 16            & 16           & 16    \\
MMAR        & 16            & 16           & 16    \\
\midrule
AlpacaEval & 128           & 32           & 128   \\
CommonEval & 128           & 32           & 128   \\
SD-QA      & 128           & 32           & 128   \\
MMSU*       & 16            & 16           & 16    \\
OBQA       & 16            & 16           & 16    \\
IFEval     & 256           & 32           & 256   \\
AdvBench   & 128           & 32           & 128   \\
\bottomrule
\end{tabular}
}

\end{table}

\setcounter{figure}{0}  
\setcounter{table}{0}   
\renewcommand{\thefigure}{C.\arabic{figure}}
\renewcommand{\thetable}{C.\arabic{table}}

\section{Additional Experiments}
\label{appendix:addition_exp}
\begin{table*}[!t]
\centering
\caption{Performance breakdown on VoiceBench. Metrics cover diverse direct QA and alignment tasks. Note that MMSU$^\ast$ in VoiceBench is derived from MMLU-Pro, which differs from the MMSU benchmark.}
\label{tab:voicebench}
\setlength{\tabcolsep}{3pt}
\renewcommand{\arraystretch}{1.00}
\resizebox{1.0\linewidth}{!}{
\begin{tabular}{l|ccccccc|c}
\toprule
\textbf{Model}                            & \textbf{AlpacaEval} & \textbf{CommonEval} & \textbf{SD-QA} & \textbf{MMSU$^\ast$}  & \textbf{OBQA}  & \textbf{IFEval} & \textbf{AdvBench} & \textbf{Overall} \\
\midrule
GPT-4o-Audio              & \textbf{4.78}       & \textbf{4.49}       & \textbf{75.50} & \textbf{80.25} & \textbf{89.23} & \textbf{76.02}  & 98.65    & \textbf{86.43}   \\
Kimi-Audio              & 4.46       & \underline{3.97}       & \underline{63.12} & \underline{62.17} & \underline{83.52} & \underline{61.10}  & \textbf{100.00}   & \underline{76.93}   \\
Qwen-2.5-Omni & \underline{4.50} & 3.84 & 56.40 & 61.70 & 80.90 & 53.50 & \underline{99.20} & 74.04 \\
Phi-4-multimodal  & 3.81 & 3.82 & 39.78 & 42.19 & 65.93 & 45.35 & 100.00 & 63.69 \\
\rowcolor{gray!30} DIFFA-2 & 3.86 & 3.67 & 40.78 & 38.13 & 61.53 & 40.76 & 85.58 & 59.63 \\
GLM-4-Voice                 & 3.97       & 3.42       & 36.98 & 39.75 & 53.41 & 25.92  & 88.08    & 55.99   \\
DiVA                      & 3.67       & 3.54       & 57.06 & 25.76 & 25.49 & 39.16  & 98.27    & 55.70   \\
Qwen2-Audio                & 3.74       & 3.43       & 35.72 & 35.72 & 49.45 & 26.33  & 96.73    & 55.34   \\
Freeze-Omni & 4.03 & 3.46 & 53.45 & 28.14 & 30.98 & 23.40 & 97.30 & 54.72 \\
Step-Audio                  & 4.13       & 3.09       & 44.21 & 28.33 & 33.85 & 27.96  & 69.62    & 49.77   \\
\rowcolor{gray!10} DIFFA        & 3.78       & 2.96       & 34.45 & 29.57 & 35.60 & 26.56  & 76.54    & 48.22    \\
LLaMA-Omni                 & 3.70       & 3.46       & 39.69 & 25.93 & 27.47 & 14.87  & 11.35    & 37.50   \\
VITA                     & 3.38       & 2.15       & 27.94 & 25.70 & 29.01 & 22.82  & 26.73    & 34.68   \\
Slam-Omni                   & 1.90       & 1.79       & 4.16  & 26.06 & 25.27 & 13.38  & 94.23    & 33.84   \\
Mini-Omni2                & 2.32       & 2.18       & 9.31  & 24.27 & 26.59 & 11.56  & 57.50    & 31.32   \\
Moshi                     & 2.01       & 1.60       & 15.64 & 24.04 & 25.93 & 10.12  & 44.23    & 27.45   \\
\bottomrule
\end{tabular}
}

\end{table*}

VoiceBench (Table~\ref{tab:voicebench}) focuses on spoken dialogue, instruction following, and safety rather than pure audio understanding.
DIFFA-2 is trained with very limited dialogue-style audio data, and its overall score (59.63) is therefore clearly lower than heavily instruction-tuned AR omnimodels such as GPT-4o-Audio, Kimi-Audio, and Qwen2.5-Omni.
Nevertheless, DIFFA-2 remains competitive with or better than several open-source baselines (e.g., GLM-4-Voice, DiVA, Qwen2-Audio, Freeze-Omni), and improves markedly over DIFFA (48.22) across most VoiceBench metrics.
This gap between strong performance on MMSU/MMAU/MMAR and mid-range performance on VoiceBench highlights the design focus of DIFFA-2: it is optimized primarily for fine-grained audio understanding rather than large-scale conversational tuning.

\end{document}